\documentclass[onecolumn,floatfix,aps,nofootinbib,showpacs, showkeys]{revtex4-2}

\usepackage{bbold}
\usepackage[dvips]{epsfig}
\usepackage[english]{babel}
\usepackage{bbm}
\usepackage{verbatim}
\usepackage{array}
\usepackage{amsmath}
\usepackage{multirow}
\usepackage{amsfonts}
\usepackage{amssymb}
\usepackage{leading}
\usepackage[dvipsnames]{xcolor}
\usepackage{subeqnarray}
\usepackage{setspace}
\usepackage{placeins}
\usepackage{float}
\usepackage{indentfirst} 

\definecolor{myblue}{rgb}{0,0,0.8}
\usepackage[colorlinks=true
,urlcolor=myblue	
,anchorcolor=myblue
,citecolor=myblue
,filecolor=myblue
,linkcolor=myblue
,menucolor=myblue
,linktocpage=true  
]{hyperref} 

\usepackage{gensymb}

\begin{document}

\title{Singular spinors and their connection}

\author{Rodolfo Jos\'e Bueno Rogerio}\email{rodolforogerio@gmail.com}
\affiliation{Institute of Physics and Chemistry, Federal University of Itajub\'a , Itajub\'a, Minas Gerais, 37500-903, Brazil.}

\begin{abstract}
\noindent{\textbf{Abstract.}} In this paper, we analyze some properties regarding singular spinors and how they are connected. The method employed here consists of mapping the spinorial structure and also the adjoint structure. Such a mathematical device is useful to determine propagators without invoking the vacuum expected value of the quantum field time-ordered product.
\end{abstract}

\maketitle

\section{Introduction}\label{intro}

The work mainly concerns some relations between singular spinors within the context of Lounesto classification. These are new categories of spinors that were unveiled in the latter years,  providing some new candidates for dark matter such as Elko spinors \cite{mdobook} and dipole spinors \cite{dharamnewfermions,rodolfodipole}, involving very interesting new possibilities from the mathematical point of view and providing a quite interesting physics.   

The work proposes to deal with quite particular properties of some \emph{non-usual} classes of spinors, the so-called singular spinors, encompassing flag-dipole, flag-pole, and dipole spinors \cite{lounestolivro}. Such spinors are fundamentally different from the usual (regular) Dirac spinors, currently used in the Standard Model (SM) of high energy physics. The singular spinors play a central role in describing what we call by \emph{Beyond the Standard Model} physics, being a very interesting topic that deserves further investigation.

Spinors stand for a sort of building blocks within the theoretical framework of the Quantum Field Theory. In this context, the most important piece of such theory is the so-called Dirac field, which play an important role in the description of spin $1/2$ charged particle --- the electron \cite{dirac1928}. In addition, the Dirac field stands just for a small piece within the realm of spinors fields.

The well known Lounesto spinor classification is a comprehensive categorization based on the bilinear covariants (physical information) that discloses the possibility of a large variety of spinors. It comprises regular and singular spinors including Dirac, flag-dipole, Majorana, and Weyl (dipole) spinors \cite{bonorapandora}. 

The aforementioned classification stands for a geometrical classification and usually classifies spinors according to their physical information, the criterion lies on the bi-spinorial densities \cite{crawford1,crawford2},usually described by the following set: $\mathcal{S}=\{\sigma,\omega,\boldsymbol{J},\boldsymbol{K},\boldsymbol{S}\}$. 
The relativistic description of the electron, accordingly the Lounesto's interpretation, establish the following: the invariant length $\sigma=\psi^{\dag}\gamma_0\psi$, the pseudo-scalar quantity $\omega=\psi^{\dag}\gamma_0\gamma^{5}\psi$, the conserved current density, defined as $\boldsymbol{J}=\psi^{\dag}\gamma_0\gamma^{\mu}\psi\gamma_{\mu}$, the quantity $\boldsymbol{K}=\psi^{\dag}\gamma_0\gamma^{\mu}\gamma^{5}\psi\gamma_{\mu}$ is interpreted as chiral current, conserved when $m = 0$ \cite[see page 51]{peskin}, and, finally, the electromagnetic momentum density $\boldsymbol{S}=\psi^{\dag}\gamma_0i\gamma^{\mu\nu}\psi\gamma_{\mu}\wedge\gamma_{\nu}$, in which $\gamma$ stands for the Dirac matrices \cite{lounestolivro,crawford1}. Such a set of bilinear forms provide the well-known invariants, given by the Fierz-Pauli-Kofink identities. 
Having said that, when at least one between $\sigma$ or $\omega$ is not both identically zero it giving rise to what is usually called \emph{regular} spinors, when $\sigma=\omega\equiv 0$, thus, giving rise to what is called \emph{singular} spinors. Regular spinors are what contains Dirac spinors and singular spinors can be further split in three other classes; according to whether $\boldsymbol{K}$ and $\boldsymbol{S}$ are not both identically zero then one obtain the \emph{flag-dipole} spinors, or with $\boldsymbol{K}\!\equiv\!0$ giving rise to the \emph{flag-pole} spinors or yet $\boldsymbol{S}\equiv 0$ giving the \emph{dipole} spinors \cite{dharamnewfermions}. Note the above classification is exhaustive.

The physical description of spinors belonging to the Lounesto classification is quite successful, except, until now, for the singular spinors. For instance, the flag-dipole spinors belong to a (very) rare set of spinors in the literature. Its application in physics was listed in specific frameworks \cite{roldaomeert,esk,tipo4epjc,tipo4dualepjc,ferrari2017,da2010elko}. 
The Majorana spinors stand for a particular example of the flag-pole spinors and the Weyl spinors are a particular class of dipole spinors with $U(1)$ gauge symmetry \cite{ferrari2017}. Thus, the singular spinors field further allocates mass dimension one fermions.

The approach taken into account here is based on a connection among the singular spinors. As we may see, it is possible to define a general singular spinor and through some mathematical conditions, we enlarge the set of singular spinors. That is, we define a primal structure that allocates the others. Then, we will examine the possibilities to map these structures and also map the physical information of the singular spinors.

The manuscript is organized as follows: In the next section, we explore some properties of the singular spinors analyzing the physical information encoded in the phase factors. In section \ref{secaomapeamento}, we define a mathematical device to map spinors and the adjoint structure. Such a protocol evince how the physical information interchange occurs. In Section \ref{quantumfields}, we develop a general method to define propagators without the necessity to employ the time-ordered product among the quantum fields. Finally, in Sect. \ref{remarks}, we conclude. 

\section{Singular spinors properties and its consequences}
Suppose a given set of dual-helicity spinors, belonging to the singular sector of the Lounesto classification 
\begin{equation}\label{spinors}
\rho= \left[ \begin{array}{c}
\alpha\Theta\phi_{l}^{*} \\ 
\beta\phi_{l}
\end{array}\right] \quad\mbox{and}\quad \varrho= \left[ \begin{array}{c}
\varsigma\phi_{r} \\ 
\vartheta\Theta\phi_{r}^{*}
\end{array}\right],
\end{equation}
where $\alpha$, $\beta$, $\varsigma$ and $\vartheta$ $\in \mathbb{C}$ represent phase factors and $\Theta$ is the well-known \emph{Wigner time-reversal operator}, which in the spin-$1/2$ representation read
\begin{equation}\label{thetawigner}
\Theta = \left[\begin{array}{cc}
0 & -1 \\ 
1 & \;\;0
\end{array} \right].
\end{equation} 
A quick inspection of the eigenstates of the helicity operator ($\vec{\sigma}\cdot\hat{p}\; \phi^{\pm} = \pm \phi^{\pm}$), in which the unit vector in spherical coordinate reads $\hat{p}=(\sin(\theta) \cos(\phi), \sin(\theta)\sin(\phi), \cos(\theta))$, directly provide 
\begin{equation}\label{comp-mais}
\phi^{+} = \sqrt{m}\left[\begin{array}{c}
\cos(\theta/2)e^{-i\phi/2} \\ 
\sin(\theta/2)e^{i\phi/2}
\end{array}\right], \quad  \phi^{-} = \sqrt{m}\left[\begin{array}{c}
-\sin(\theta/2)e^{-i\phi/2} \\ 
\cos(\theta/2)e^{i\phi/2}
\end{array}\right],
\end{equation} 
carrying positive helicity and the negative helicity, respectively.
The operator introduced in Eq.\eqref{thetawigner} combined with the components \eqref{comp-mais}, allow to define the following relations
\begin{equation}\label{compsviatheta}
\Theta\phi^{*\;+} = \phi^{-}, \;\;  \Theta\phi^{*\;-} = -\phi^{+}.
\end{equation}
With $\Theta\phi^{*\;+}$ and $\Theta\phi^{*\;-}$ behaving as negative and positive helicity components, respectively. 
To make the notation clear, we purposely omitted the Lorentz boost factors and also the helicity indexes. 
Both spinors in \eqref{spinors} carry the same physical information, therefore, for the sake of simplicity, all the calculations will be performed taking into account only the $\rho$ spinor. 
We emphasize when the phase factors are constrained to the following condition: $|\alpha|^2\neq|\beta|^2$, then, we are handling the most general dual-helicity spinor, namely flag-dipole spinor.
Now, if one impose to $\rho$ to hold conjugacy under charge conjugation operator, $\mathcal{C}=\gamma_2 \mathcal{K}$, in which $\mathcal{K}$ stands for the algebraic complex conjugation, thus, it yields
\begin{equation}\label{eq5}
\mathcal{C}\rho^{\pm} = \pm \rho^{\pm},
\end{equation}
where the upper indexes ``$\pm$'' are related with the associated eigenvalues. Such constraint force the $\rho$ spinor to be written as it follows
\begin{equation}\label{conjugados}
\rho^{+} = \left[ \begin{array}{c}
i\beta^{*}\Theta\phi_{l}^{*} \\ 
\beta\phi_l
\end{array} \right] \quad\mbox{and}\quad \rho^{-} = \left[ \begin{array}{c}
-i\beta^{*}\Theta\phi_{l}^{*} \\ 
\beta\phi_l
\end{array} \right].
\end{equation}
Note if one set $\beta=1$, it matches with the Elko spinors described in \cite{mdobook}. The dual-helicity spinors in \eqref{conjugados} stand for a specific sub-class of the flag-pole spinors, as previously observed in \cite{rodolfoconstraints}. The simple fact of imposing to a flag-dipole spinor hold conjugacy under charge conjugation operator,  lead to a flag-pole spinor, restricting, thus, the degree of freedom of the phase factors. Such an imposition force the phase factors present in both components to have equal moduli and, thus, the $\boldsymbol{K}$ bilinear form is identically null, leading, then, the flag-dipole set of bilinear forms $\mathcal{S}_{f.d}=\{0,0,\boldsymbol{J},\boldsymbol{K},\boldsymbol{S}\}$ to $\mathcal{S}_{f.p}=\{0,0,\boldsymbol{J},0,\boldsymbol{S}\}$. 

Another interesting fact that we bring to the scene follows: if one set $\alpha=0$ or $\beta=0$ in \eqref{spinors}, it becomes clear one directly obtain dipole spinors, which quite recently was shown to stand for a new set of mass-dimension-one fermion, being also strong candidates to describe dark matter \cite{dharamnewfermions}. Thus, the dipole spinors can be displayed in the following fashion
\begin{equation}
\rho_{dipole_{1}} = \left[ \begin{array}{c}
\alpha\Theta\phi^{*} \\ 
0
\end{array} \right] \quad\mbox{and}\quad \rho_{dipole_{2}} = \left[ \begin{array}{c}
0 \\ 
\beta\Theta\phi^{*}
\end{array} \right].
\end{equation}
Notice the last phase condition yield $\mathcal{S}_{f.d}\rightarrow \mathcal{S}_{dipole} = \{0,0,\boldsymbol{J},\boldsymbol{K},0\}$. Thus, one might infer that most of the physical information contained therein is exclusively represented by the phase factors and not only by the spinor itself.
Moreover, we highlight that such a task can not be accomplished by a flag-pole spinor \eqref{conjugados}, since flag-pole spinor has a (very) restricted degree of freedom, carrying only one phase factor ($\beta$), and imposing $\beta=0$ in eq.\eqref{conjugados}, we obtain a non-physical instance. Leading to conclude that dipole spinors can not be directly experienced from flag-pole spinors.    

We also emphasize that the protocol we are exploring seems to be unique. We start with a general structure, a flag-dipole spinor, and, under certain requirements, we transmute to a flag-pole spinor or to transmute from a flag-dipole spinor to a dipole spinor, however, the opposite does not hold. It is impossible to construct any other spinor from only one Weyl spinor \cite{rodolfosubliminal}. The previous results are in agreement with \cite{chengflagdipole}.

\section{Connecting the physical information}\label{secaomapeamento}
Bearing in mind a huge variety of spinors in the $QFT$ context, as shown in \cite{mdobook,dharamnewfermions}, in this section we derive a mapping program based on the existing well-known spinorial structures. Such an approach aims to connect the spinor's physical information and turn possible to understand how they are connected.   

Looking towards defining a connection among different spinorial classes, all the analysis developed here is based on the fundamental matters: \emph{a)} first, we map $\psi$ into $\Phi$, one holding the Dirac adjoint structure and the other a more involved and general structure, \emph{b)} finally, we take both spinors carrying a more general adjoint structure. 
Thus, such a task is fully accomplished by defining a mapping protocol between spinors belonging to distinct classes. 

Let $\psi$ and $\Phi$ two distinct singular spinors 
\begin{equation}\label{map}
\psi=\mathcal{M}\Phi,
\end{equation}
in which $\mathcal{M}$ stand for a $4\times 4$ diagonal matrix, indeed $\mathcal{M}^{-1}$ exists ensuring an invertible map. The last statements translate into \footnote{In general grounds, the anti-particle counterpart is defined under the following replacement $\epsilon_1\rightarrow -\epsilon_1$ or $\epsilon_2\rightarrow -\epsilon_2$.}
\begin{eqnarray}
\psi = \left(\begin{array}{cc}
\epsilon_{1}\mathbbm{1}_{(2\times 2)} & 0 \\ 
0 & \epsilon_{2}\mathbbm{1}_{(2\times 2)}
\end{array} \right)\left( \begin{array}{c}
\alpha \Theta\phi^{*} \\ 
\beta \phi
\end{array}  \right).
\end{eqnarray}
A careful inspection of the parameters $\epsilon_1$ and $\epsilon_2$ open doors to connect (in the most general way) singular spinors, as presented in the table below
\begin{table}[htbp]
\centering
\begin{tabular}{cc|c|c|c}
\hline 
 & \;\;\;\;\;\;\;\;\;\;$\epsilon_{1}$\;\;\;\;\;\;\;\;\;\;\;\; & \;\;\;\;\;\;\;\;\;\;\;\;$\epsilon_{2}$\;\;\;\;\;\;\;\;\;\; & \;\;\;\;$\psi$ \emph{belong to the following Lounesto class}\;\;\;\; & \;\;\;\;\emph{Constraints among the phases}\;\;\;\;\\ 
\hline 
\hline 
 & $\forall\; {\rm I\!R}$ & $\forall\; {\rm I\!R}$ & 4 & $|\epsilon_{1}\alpha|^2\neq|\epsilon_{2}\beta|^2$\\
   & $\forall\; {\rm I\!R}$ & $\forall\; {\rm I\!R}$ & 5 & $|\epsilon_{1}\alpha|^2=|\epsilon_{2}\beta|^2$\\
 & $\forall\;\mathbb{C}$ & $\forall\;\mathbb{C}$ & 4 & $|\epsilon_{1}\alpha|^2\neq|\epsilon_{2}\beta|^2$\\ 
  & $\forall\;\mathbb{C}$ & $\forall\;\mathbb{C}$ & 5 & $|\epsilon_{1}\alpha|^2=|\epsilon_{2}\beta|^2$\\ 
  & $\forall\;{\rm Im}$ & $\forall\;{\rm Im}$ & 4 &$|\epsilon_{1}\alpha|^2\neq|\epsilon_{2}\beta|^2$\\ 
 & $\forall\;{\rm Im}$ & $\forall\;{\rm Im}$ & 5 &$|\epsilon_{1}\alpha|^2=|\epsilon_{2}\beta|^2$\\  
 & $\alpha^*$ & $\beta^*$ & 4 & $|\alpha|^2\neq|\beta|^2$ \\ 
  & $\alpha^*$ & $\beta^*$ & 5 & $|\alpha|^2=\beta|^2$ \\
    & $ix/\alpha$ & $iy/\beta$ & 4 & $|x|^2\neq|y|^2$\\    
    & $ix/\alpha$ & $iy/\beta$ & 5 & $|x|^2=|y|^2$\\ 
 & $x/\alpha$ & $y/\beta$ & 4 & $|x|^2\neq|y|^2$\\  
  & $x/\alpha$ & $y/\beta$ & 5 & $|x|^2=|y|^2$\\ 
  & 1 & $\alpha/\beta$ & 5 & - \\ 
 & $\beta/\alpha$ & 1 & 5 &-\\ 
 & $ix/\alpha$ & $y/\beta$ & 4 & $|x|^2\neq|y|^2$\\  
  & $ix/\alpha$ & $y/\beta$ & 5 & $|x|^2=|y|^2$\\ 
  & 0 & $\forall\;\mathbb{C}$, $\forall\; {\rm I\!R}$,$\forall\;{\rm Im}$   & 6 & \emph{invertible map not allowed} \\ 
  & $\forall\;\mathbb{C}$, $\forall\; {\rm I\!R}$,$\forall\;{\rm Im}$ & 0 & 6 &\emph{invertible map not allowed}\\ 
\hline 
\hline
\end{tabular}
\label{tabela1}
\caption{The phases possibilities to connect dual-helicity spinors and classify them within Lounesto classification. Here we examined all the possibilities, any other may be written in terms of the previously defined above. Notice that $x$ and $y$ parameters stand for numbers that depend on the kind of $\alpha$ and $\beta$ phases. For instance, if $\alpha$ and $\beta$ $\in\rm I\!R$, thus, of course $x$ and $y$ $\in\rm I\!R$.} 
\end{table}
\newpage
\subsection{On the first mapping program}\label{secao1mapeamento}
We start the algorithm by imposing the following 
\begin{eqnarray}
\bar{\psi} = \psi^{\dag}\gamma_0,\quad\mbox{and}\quad \stackrel{\neg}{\Phi} = [\Xi_{1}\Phi]^{\dag}\gamma_0,
\end{eqnarray}
where $\Xi_{1} \neq \mathbbm{1}$, $\Xi_{1}^2\propto \mathbbm{1}$ and $\Xi_{1}^{-1}$ indeed exists. Taking into account Eq.\eqref{map}, one is able to write the mapped adjoint structure in the following fashion
\begin{equation}\label{dualdiracdualnovo}
\bar{\psi} = \stackrel{\neg}{\Phi}\gamma_0\Xi_{1}^{\dag}\mathcal{M}^{\dag}\gamma_0,
\end{equation}
and
\begin{equation}\label{dualnovodualdirac}
\stackrel{\neg}{\Phi} = \bar{\psi}\gamma_0{(\Xi_{1}^{\dag}\mathcal{M}^{\dag})}^{-1}\gamma_0.
\end{equation}
Notice that equations \eqref{dualdiracdualnovo} and \eqref{dualnovodualdirac} bring to scene a strong relation among both adjoint structures --- turning explicit how one can define a relation among distinct objects.

Now, replacing $\bar{\psi}$ by $\stackrel{\neg}{\Phi}\gamma_0\Xi_{1}^{\dag}\mathcal{M}^{\dag}\gamma_0$ and $\psi$ by $\mathcal{M}\Phi$ in the Lounesto classification, it allows to re-write the bilinear forms as follow
\begin{eqnarray}\label{covariantes1}
\tilde{\sigma}_{\psi} =\stackrel{\neg}{\Phi}\gamma_0\Xi_{1}^{\dag}\mathcal{M}^{\dag}\gamma_0\mathcal{M}\Phi,\; 
\\
\tilde{\omega}_{\psi} = i\stackrel{\neg}{\Phi}\gamma_0\Xi_{1}^{\dag}\mathcal{M}^{\dag}\gamma_0\gamma_5\mathcal{M}\Phi, 
\\  
\tilde{\mathbf{J}}_{\psi} = \stackrel{\neg}{\Phi}\gamma_0\Xi_{1}^{\dag}\mathcal{M}^{\dag}\gamma_0 \gamma_\mu \mathcal{M}\Phi \gamma^\mu, 
\\
\tilde{\mathbf{K}}_{\psi}= -\stackrel{\neg}{\Phi}\gamma_0\Xi_{1}^{\dag}\mathcal{M}^{\dag}\gamma_0 \gamma_{5} \gamma_\mu \mathcal{M}\Phi \gamma^\mu,  \\ 
\tilde{\mathbf{S}}_{\psi}=  \stackrel{\neg}{\Phi}\gamma_0\Xi_{1}^{\dag}\mathcal{M}^{\dag}\gamma_0 i \gamma_{\mu}\gamma_{\nu}\mathcal{M}\Phi \gamma^\mu \wedge \gamma^\nu. 
\end{eqnarray}   
We highlight although all the bilinear forms above are written in terms of $\Phi$, however, it holds information about the $\psi$ spinor. Bearing in mind the elements of the usual Clifford algebra stand for 
\begin{equation}\label{cliffordbasis}
\Gamma = \lbrace \mathbbm{1}, \gamma_{\mu}, \gamma_5, \gamma_5\gamma_{\mu}, \gamma_{\mu}\gamma_{\nu}\rbrace, 
\end{equation}  
after performing the procedure given in Eq.\eqref{dualdiracdualnovo} and Eq.\eqref{dualnovodualdirac}, the usual elements of the Clifford algebra in Eq.\eqref{cliffordbasis}  now can be written in its ``mapped'' form
\begin{equation}
\Gamma \rightarrow \tilde{\Gamma} = \gamma_0\Xi_{1}^{\dag}\mathcal{M}^{\dag}\gamma_0\Gamma\mathcal{M},
\end{equation}
evincing a close relation among $\Gamma$ altogether with the deformed one $\tilde{\Gamma}$.

\subsection{On the second mapping program}\label{secao2mapeamento}
Now we deal with a more involved pattern. From now on, we assume to both spinors, $\psi$, and $\Phi$ to carry a more robust adjoint structure, thus, we define
\begin{eqnarray}\label{map2}
\stackrel{\neg}{\Phi} = [\Xi_{1}\Phi]^{\dag}\gamma_0 \quad\mbox{and}\quad \stackrel{\neg}{\psi} = [\Xi_{2}\psi]^{\dag}\gamma_0, 
\end{eqnarray}
such relations allow one to define the mapped adjoint structure as it follows
\begin{equation}
\stackrel{\neg}{\Phi} = \stackrel{\neg}{\psi}\gamma_0{(\Xi_{1}^{\dag}\mathcal{M}^{\dag}\Xi_{2}^{\dag})}^{-1}\gamma_0,
\end{equation}
and 
\begin{equation}
\stackrel{\neg}{\psi} = \stackrel{\neg}{\Phi}\gamma_0\Xi_{1}^{\dag}\mathcal{M}^{\dag}\Xi_{2}^{\dag}\gamma_0.
\end{equation}
    
With the last set of equations at hands, we may write the following
\begin{eqnarray}\label{covariantes2}
\tilde{\sigma}_{\psi} =\stackrel{\neg}{\Phi}\gamma_0\Xi_{1}^{\dag}\mathcal{M}^{\dag}\Xi_{2}^{\dag}\gamma_0\mathcal{M}\Phi,
\\
 \tilde{\omega}_{\psi} = i\stackrel{\neg}{\Phi}\gamma_0\Xi_{1}^{\dag}\mathcal{M}^{\dag}\Xi_{2}^{\dag}\gamma_0\gamma_5\mathcal{M}\Phi, 
 \\  
\tilde{\mathbf{J}}_{\psi} = \stackrel{\neg}{\Phi}\gamma_0\Xi_{1}^{\dag}\mathcal{M}^{\dag}\Xi_{2}^{\dag}\gamma_0 \gamma_\mu \mathcal{M}\Phi \gamma^\mu, 
\\
 \tilde{\mathbf{K}}_{\psi}= -\stackrel{\neg}{\Phi}\gamma_0\Xi_{1}^{\dag}\mathcal{M}^{\dag}\Xi_{2}^{\dag}\gamma_0 \gamma_{5} \gamma_\mu \mathcal{M}\Phi \gamma^\mu,  
 \\ 
\tilde{\mathbf{S}}_{\psi}=  \stackrel{\neg}{\Phi}\gamma_0\Xi_{1}^{\dag}\mathcal{M}^{\dag}\Xi_{2}^{\dag}\gamma_0 i \gamma_{\mu} \gamma_{\nu} \mathcal{M}\Phi \gamma^\mu \wedge \gamma^\nu. 
\end{eqnarray}
notice the above bilinear forms hold information concerning the $\psi$ spinor. The presence of the $\mathcal{M}$, $\Xi_1$ and $\Xi_2$ operators lead to
\begin{equation}
\Gamma \rightarrow \tilde{\Gamma} =   \gamma_0\Xi_{1}^{\dag}\mathcal{M}^{\dag}\Xi_{2}^{\dag}\gamma_0\Gamma\mathcal{M}.
\end{equation}
The results above may be summarized as follows: when mapping spinors all the physical information is close related accordingly to the way that they are connected: via $\mathcal{M}$ and $\Xi$ operators.
\subsubsection{Extending the programme}\label{dualfinal}
Here we investigate a more specific and a (very) very particular framework. Taking into account the adjoint structure previously defined in \cite{tipo4epjc, aaca}, thus, we may set the following relations
\begin{eqnarray}\label{ultimaformadual}
\stackrel{\neg}{\Phi} = [\Xi_{1}\Phi]^{\dag}\gamma_0\mathcal{O}_{\Phi} \quad\mbox{and}\quad \stackrel{\neg}{\psi} = [\Xi_{2}\psi]^{\dag}\gamma_0\mathcal{O}_{\psi}, 
\end{eqnarray}
in which the $\mathcal{O}$ operator brings information about the spin sums. So, the adjoint structures defined in Eq.\eqref{ultimaformadual}, allow to define
\begin{equation}
\stackrel{\neg}{\Phi} = \stackrel{\neg}{\psi}\mathcal{O}^{-1}_{\psi}\gamma_0{(\Xi_{1}^{\dag}\mathcal{M}^{\dag}\Xi_{2}^{\dag})}^{-1}\gamma_0\mathcal{O}_{\Phi},
\end{equation}
and 
\begin{equation}
\stackrel{\neg}{\psi} = \stackrel{\neg}{\Phi}\mathcal{O}^{-1}_{\Phi}\gamma_0\Xi_{1}^{\dag}\mathcal{M}^{\dag}\Xi_{2}^{\dag}\gamma_0\mathcal{O}_{\psi}.
\end{equation}
So far, the above dual structures refer to the most complex examples ever introduced in the literature.

\section{The singular spinor propagators connection}\label{quantumfields}
This section is reserved to define an algorithm to connect propagators for singular spinors, taking into account the previous definitions in the sections \ref{secao1mapeamento} and \eqref{secao2mapeamento}.
Before defining the most general propagator for the singular spinors, we might first make explicit the quantum field operators. That said, it reads
\begin{eqnarray}
\mathcal{F}(x)\!=\!\int\frac{d^3 p}{(2\pi)^3}\frac{1}{\sqrt{2E(\boldsymbol{p})}}\sum_{spin}\Big[c_{k}(\boldsymbol{p})\psi_{k}(\boldsymbol{p}) e^{-ip_{\mu}x^{\mu}}
+d_{k}^{\dag}(\boldsymbol{p})\chi_{k}(\boldsymbol{p}) e^{ip_{\mu}x^{\mu}}\Big],
\end{eqnarray}
whereas the adjoint field operator is given by
\begin{eqnarray}
\bar{\mathcal{F}}(x)\!=\!\int\frac{d^3 p}{(2\pi)^3}\frac{1}{\sqrt{2E(\boldsymbol{p})}}\sum_{spin}\Big[c_{k}^{\dag}(\boldsymbol{p})\bar{\psi}_{k}(\boldsymbol{p}) e^{ip_{\mu}x^{\mu}}
+d_{k}(\boldsymbol{p})\bar{\chi}_{k}(\boldsymbol{p}) e^{-ip_{\mu}x^{\mu}}\Big],
\end{eqnarray}
in which $\chi$ stand for the anti-particle spinor. The creation and annihilation operators, $c^{\dag}(\boldsymbol{p})$ and $c(\boldsymbol{p})$, yield the usual fermionic relations
\begin{eqnarray}
&&\{c_{k}(\boldsymbol{p}),c_{k^{\prime}}^{\prime\dag}(\boldsymbol{p}^{\prime})\}=\{d_{k}(\boldsymbol{p}),d_{k^{\prime}}^{\prime\dag}(\boldsymbol{p}^{\prime})\}=(2\pi)^3\delta^{3}(\boldsymbol{p}-\boldsymbol{p}^{\prime})\delta_{kk^{\prime}},
\\
&&\{c_{k}(\boldsymbol{p}),d_{k^{\prime}}(\boldsymbol{p}^{\prime})\}=\{c^{\dag}_{k}(\boldsymbol{p}),d^{\dag}_{k^{\prime}}(\boldsymbol{p}^{\prime})\}=0.
\end{eqnarray}  
Thus, the Feynman-Dyson propagator is given by a time-ordered product of $\mathcal{F}(x)$ and $\bar{\mathcal{F}}(x)$ functions
\begin{eqnarray}\label{fdpropagator}
i\mathcal{D}(x-x^{\prime})=\langle 0 \vert \mathcal{F}(x)\bar{\mathcal{F}}(x^{\prime}) \vert 0 \rangle  \theta(x^{0}-x^{\prime 0})- \langle 0 \vert \bar{\mathcal{F}}(x^{\prime})\mathcal{F}(x) \vert 0 \rangle \theta(x^{\prime 0}-x^{0}),
\end{eqnarray}
where the Heaviside function, $\theta(t)$, read
\begin{eqnarray}
\theta(t)\!=\!-\frac{1}{2\pi i}\int_{-\infty}^{\infty} ds\; \frac{e^{-ist}}{s+i\epsilon}.
\end{eqnarray}   
Now, having defined the quantum field operators, we proceed evaluating term by term of \eqref{fdpropagator}, then, we easily obtain
\begin{eqnarray}\label{fd1}
&&\langle 0\vert \mathcal{F}(x)\bar{\mathcal{F}}(x^{\prime}) \vert 0 \rangle\theta(x^{0}-x^{\prime 0})\!=\! \nonumber\\
&&-\frac{1}{2\pi i}\int\frac{d^3 p}{(2\pi)^3}\frac{1}{2E(\boldsymbol{p})}
 \int_{-\infty}^{\infty}ds\sum_{spin}\psi_{k}(\boldsymbol{p})\bar{\psi}_{k}(\boldsymbol{p})\frac{e^{-i(s+p_{0})(x^{0}-x^{\prime 0})+i\boldsymbol{p}(\boldsymbol{x}-\boldsymbol{x}^{\prime})}}{s+i\epsilon},
\end{eqnarray}
under the assumption $x^0> x^{\prime0}$, and 
\begin{eqnarray}\label{fd2}
&&\langle 0 \vert \bar{\mathcal{F}}(x^{\prime}){\mathcal{F}}(x) \vert 0 \rangle \theta(x^{\prime 0}-x^{0})\!=\! \nonumber\\
&&-\frac{1}{2\pi i}\int\frac{d^3 p}{(2\pi)^3}\frac{1}{2E(\boldsymbol{p})}\int_{-\infty}^{\infty}ds\sum_{spin}\chi_{k}(\boldsymbol{p})\bar{\chi}_{k}(\boldsymbol{p})\frac{e^{i(s+p_{0})(x^{0}-x^{\prime 0})-i\boldsymbol{p}(\boldsymbol{x}-\boldsymbol{x}^{\prime})}}{s+i\epsilon},
\end{eqnarray}
for $x^0< x^{\prime0}$. 
Finally, inserting Eq.\eqref{fd1} and Eq.\eqref{fd2} into Eq.\eqref{fdpropagator} and using some of Heaviside function properties, one obtain the general form of the Feynman-Dyson propagator, furnishing
\begin{eqnarray}\label{fd3}
\mathcal{D}(x-x^{\prime})&=&\int\frac{d^4 p}{(2\pi)^4}\frac{1}{2E(\boldsymbol{p})}\Bigg[\frac{\sum_{spin}\psi_{k}(\boldsymbol{p})\bar{\psi}_{k}(\boldsymbol{p})(p_0+\sqrt{p_{j}p^{j}+m^2})}{p^{2}-m^2+i\epsilon}
\nonumber\\
&+& \frac{\sum_{spin}\chi_{k}(-\boldsymbol{p})\bar{\chi}_{k}(-\boldsymbol{p})(p_0-\sqrt{p_{j}p^{j}+m^2})}{p^2-m^2+i\epsilon} \Bigg]e^{-ip_{\mu}(x^{\mu}-x^{\prime\mu})},
\end{eqnarray}
where $j=1, 2, 3$. Essentially, the propagator in \eqref{fd3} can be written as 
\begin{eqnarray}
\mathcal{D}(x-x^{\prime}) = \mathcal{D}(x-x^{\prime})_{particle}+\mathcal{D}(x-x^{\prime})_{anti-particle}.
\end{eqnarray} 
So far we have defined the most general propagator. Now, we employ the definitions derived in section \ref{secao2mapeamento} to re-write the spin-sums of the above propagator. Thus, after a bit of algebra, one obtain
\begin{eqnarray}\label{propagadormapeado}
\mathcal{D}_{\mathcal{M},\Xi}(x-x^{\prime}) = \mathcal{M}\mathcal{D}(x-x^{\prime})\gamma_0\Xi_1^{\dag}\mathcal{M}^{\dag}\Xi_2^{\dag}\gamma_0.
\end{eqnarray}   
In general grounds, the main result encoded in the propagator above is: once we have defined the quantity $\mathcal{D}(x-x^{\prime})$ for a specific singular spinor, the $\mathcal{D}_{\mathcal{M},\Xi}(x-x^{\prime})$ is essentially given by a product of the operators $\mathcal{M}$ and $\gamma_0\Xi_1^{\dag}\mathcal{M}^{\dag}\Xi_2^{\dag}\gamma_0$, Allowing, thus, to fully define the propagator for other singular spinors. We remark that the above approach take into account examples where $\Xi$ operator stand the same for both particle and antiparticle spinor (as developed in Ref \cite{mdobook}), nonetheless, if is not the case (as in Ref \cite{dharamnewfermions}), then Eq.\eqref{propagadormapeado} should be split in a sum of two propagators, one holding all the information related to the particle spinor and the other holding information of the antiparticle spinor, given by
 \begin{eqnarray}\label{propagadormapeado2}
\mathcal{D}_{\mathcal{M},\Xi}(x-x^{\prime}) = [\mathcal{M}\mathcal{D}(x-x^{\prime})\gamma_0\Xi_1^{\dag}\mathcal{M}^{\dag}\Xi_2^{\dag}\gamma_0]_{_{particle}}+[\mathcal{M}\mathcal{D}(x-x^{\prime})\gamma_0\Xi_1^{\dag}\mathcal{M}^{\dag}\Xi_2^{\dag}\gamma_0]_{_{anti-particle}}.
\end{eqnarray}     
Nonetheless, if one wishes to deal with the adjoint structure previously defined in Sect. \ref{dualfinal}, a more involved propagator emerge
 \begin{eqnarray}
\mathcal{D}_{\mathcal{M},\Xi}(x-x^{\prime}) &=& [\mathcal{M}\mathcal{D}(x-x^{\prime})\mathcal{O}^{-1}_{\Phi}\gamma_0\Xi_1^{\dag}\mathcal{M}^{\dag}\Xi_2^{\dag}\gamma_0\mathcal{O}_{\psi}]_{_{particle}}\nonumber\\
&+&[\mathcal{M}\mathcal{D}(x-x^{\prime})\mathcal{O}^{-1}_{\Phi}\gamma_0\Xi_1^{\dag}\mathcal{M}^{\dag}\Xi_2^{\dag}\gamma_0\mathcal{O}_{\psi}]_{_{anti-particle}}.
\end{eqnarray}
Notice the general propagator definition $\mathcal{D}_{\mathcal{M},\Xi}(x-x^{\prime})$, is accomplished by a matrix product.

\subsection{A guide to the existing propagators: An explicit example}
So far we algebraically determine how propagators may be connected. Now, we may show the usefulness of the above prescription. To illustrate the method, we define the propagator defined in \cite{dharamnewfermions} in terms of the Elko propagator \cite{jcap}. Suppose, then, the quantity in \eqref{propagadormapeado2} written in the following fashion
\begin{eqnarray}
\mathcal{D}_{dipole}(x-x^{\prime}) &=& [\mathcal{M}\mathcal{D}_{Elko}(x-x^{\prime})\gamma_0\Xi_{Elko}^{\dag}\mathcal{M}^{\dag}\Xi_{dipole}^{\dag}\gamma_0]_{_{particle}}\nonumber\\
&+&[\mathcal{M}\mathcal{D}_{Elko}(x-x^{\prime})\gamma_0\Xi_{Elko}^{\dag}\mathcal{M}^{\dag}\Xi_{dipole}^{\dag}\gamma_0]_{_{anti-particle}}, 
\end{eqnarray}  
note the last equation above provide the $\mathcal{D}_{dipole}(x-x^{\prime})$ based on the Elko's propagator. Remembering the following properties: $\Xi_{Elko}=m^{-1}\mathcal{G}(\phi)\gamma_{\mu}p^{\mu}$ \cite{aaca}, $\Xi_{dipole}=m^{-1}\gamma_{\mu}p^{\mu}$ for particle and $\Xi_{dipole}=-m^{-1}\gamma_{\mu}p^{\mu}$ for anti-particle \cite{dharamnewfermions}, the phases presented in Table 1 and from Eq.\eqref{map} we have $\mathcal{M}= diag(1, 1, 0, 0)$ for particles and $\mathcal{M}= diag(0, 0, 1, 1)$ for anti-particle. Thus,to coherently illustrate the algorithm presented in Eq.\eqref{map2}, the Elko propagator is given by \cite{jcap}
\begin{eqnarray}
\mathcal{D}_{Elko}(x-x^{\prime})&=&\int\frac{d^4 p}{(2\pi)^4}\frac{1}{2E(\boldsymbol{p})}\Bigg[\frac{\sum_{h}\lambda^S_{h}(\boldsymbol{p})\stackrel{\neg}{\lambda}^{S}_{h}(\boldsymbol{p})(p_0+\sqrt{p_{j}p^{j}+m^2})}{p^{2}-m^2+i\epsilon}\nonumber\\
&+&\frac{\sum_{h}\lambda^{A}_{h}(-\boldsymbol{p})\stackrel{\neg}{\lambda}^{A}_{h}(-\boldsymbol{p})(p_0-\sqrt{p_{j}p^{j}+m^2})}{p^2-m^2+i\epsilon} \Bigg]e^{-ip_{\mu}(x^{\mu}-x^{\prime\mu})},
\end{eqnarray}      
thus, accordingly with \eqref{propagadormapeado2}, we obtain
\begin{eqnarray}\label{666}
\mathcal{D}_{dipole}(x-x^{\prime})= \Bigg[\mathcal{M}\int\frac{d^4 p}{(2\pi)^4}\frac{1}{2E(\boldsymbol{p})}\Bigg(\frac{\sum_{h}\lambda^S_{h}(\boldsymbol{p})\stackrel{\neg}{\lambda}^{S}_{h}(\boldsymbol{p})(p_0+\sqrt{p_{j}p^{j}+m^2})}{p^{2}-m^2+i\epsilon}
\Bigg)e^{-ip_{\mu}(x^{\mu}-x^{\prime\mu})}\Delta\Bigg]_{_{particle}}\nonumber\\
+\Bigg[\mathcal{M}\int\frac{d^4 p}{(2\pi)^4}\frac{1}{2E(\boldsymbol{p})}\Bigg(\frac{\sum_{h}\lambda^{A}_{h}(-\boldsymbol{p})\stackrel{\neg}{\lambda}^{A}_{h}(-\boldsymbol{p})(p_0-\sqrt{p_{j}p^{j}+m^2})}{p^2-m^2+i\epsilon} \Bigg)e^{-ip_{\mu}(x^{\mu}-x^{\prime\mu})}\Delta\Bigg]_{_{anti-particle}}, 
\end{eqnarray}
in which we have defined $\Delta=\gamma_0\Xi_{Elko}^{\dag}\mathcal{M}^{\dag}\Xi_{dipole}^{\dag}\gamma_0$, thus, we have
\begin{eqnarray}
\Delta_{particle} = \left(\begin{array}{cccc}
0 & 0 & 0 & 0 \\ 
0 & 0 & 0 & 0 \\ 
0 & -ie^{-i\phi} & 0 & 0 \\ 
ie^{i\phi} & 0 & 0 & 0
\end{array}\right), 
\end{eqnarray}
and
\begin{eqnarray}
\Delta_{anti-particle} = \left(\begin{array}{cccc}
0 & 0 & 0 & -ie^{-i\phi} \\ 
0 & 0 & ie^{i\phi} & 0 \\ 
0 & 0 & 0 & 0 \\ 
0 & 0 & 0 & 0
\end{array}\right).
\end{eqnarray}
For the sake of simplicity, we take into account the approach given in Sect.\ref{secao2mapeamento}, where the spin sums in Eq.\eqref{666} carry the $\mathcal{G}(\phi)$ operator \cite{jcap}. Otherwise, one must employ the program developed in \ref{dualfinal}, which would make the above calculations more complicated, however, furnishing the desired result.

Having said that, performing the calculations in Eq.\eqref{666} we are able to compute $\mathcal{D}_{dipole}(x-x^{\prime})$, yielding
\begin{eqnarray}
\mathcal{D}_{dipole}(x-x^{\prime}) = \int\frac{d^4 p}{(2\pi)^4}e^{-ip_{\mu}(x^{\mu}-x^{\prime\mu})}\frac{\mathbbm{1}}{p_{\mu}p^{\mu}-m^2+i\epsilon}.
\end{eqnarray}
In agreement with the propagator defined in \cite{dharamnewfermions}. 

We emphasize that the very same approach may be taken for any other singular spinor. What we have developed is a simple demonstration of how the method works and can be applied. Since we previously have the dual structure of a given singular spinor and the $\mathcal{M}$ mapping matrix, we can simply connect propagators, without the necessity of computing the vacuum expected value of a time-ordered product of quantum fields.

\section{Concluding remarks}\label{remarks}

The comprehension of unorthodox classes of spinors is an interesting path in mathematical and high energy physics, and this paper includes some aspects regarding one of these lesser know classes of spinors. 
As such singular spinors may be used to build a physically meaningful theory once we can understand how flag-pole and dipole spinors may be defined from flag-dipole spinors. 

In this report we show the possibility of, under certain physical and mathematical requirements, transmuting from a flag-dipole spinor to a flag-pole spinor or to transmute to dipole spinors. Therefore, as commented along with the paper, it is impossible to obtain dipole spinors from a flag-pole spinor given its restricted degree of freedom. Thus, we show that flag-dipole spinors can be understood as primitive structures within the singular sector of the Lounesto classification.
It is worth mentioning that all the relevant physical information is encoded in the phase factors.   

Going further, we reported a close connection among singular spinors, presented in a mapping program. We have shown that it is possible to obtain physical information regarding one particular spinor, however, written in terms of another, although, the price to be paid is the Clifford algebra deformation. As can be seen, this deformation is evident in the bilinear forms and the quantum field propagator. Finally, to illustrate the algorithm developed here, we explicitly show how it is possible to ascertain the propagator of two distinct theories --- for instance, Elko and dipole spinors, however, as can be seen, the developed mechanism is not restricted to just these two examples.

\bibliographystyle{unsrt}
\bibliography{refs}

\end{document}